\documentclass{jps-cp}
\usepackage{color}

\title{Magnetic Properties of an Effective Spin-$\frac{1}{2}$ Triangular-Lattice Compound LiYbS$_2$}

\author{K. M. \textsc{Ranjith}$^{1,\dagger}$, Ph. \textsc{Schlender}$^{2}$, Th. \textsc{Doert}$^{2}$, and M. \textsc{Baenitz}$^{1}$}
\inst{$^{1}$Max Planck Institute for Chemical Physics of Solids, Dresden, Germany
 \\
$^{2}$Faculty of Chemistry and Food Chemistry, TU Dresden, Dresden, Germany}

\email{ranjith.kumar@cpfs.mpg.de}

\recdate{September 15, 2019}

\abst{Here, we report the synthesis and magnetic properties of a Yb-based triangular-lattice compound LiYbS$_2$. At low temperatures, it features an effective spin-$\frac{1}{2}$ state due to the combined effect of crystal electric field and spin orbit coupling. Magnetic susceptibility measurements and $^7$Li nuclear magnetic resonance experiments reveal the absence of magnetic long range ordering down to 2~K, which suggests a possible quantum spin liquid ground state. A dominant antiferromagnetic nearest neighbour exchange interaction $J/k_{\rm B}\simeq$ 5.3~K could be extracted form the magnetic susceptibility. The NMR linewidth analysis yields the coupling constant between the Li nuclei and Yb$^{3+}$ ions which was found to be purely dipolar in nature.
 }

\kword{quantum spin liquid, triangular lattice, $^7$Li NMR.}

\begin{document}
\maketitle

\section{Introduction}

Quantum spin liquid (QSL), a highly entangled state with fractionalized excitations that evade magnetic long-range order down to zero temperature, has been attracting the attention of condensed matter research for several decades~\cite{anderson1973,balents2010,Zhou2017,Wen2019}. Yb-based dichalcogenide delafossites $A$Yb$X_2$ ($A$ = monovalent ion; $X$ = divalent chalcogen ion) are recently reported as the promising candidates for realizing QSL ground states~\cite{liu2018,baenitz2018,Ranjith2019,Ding2019,Bordelon2019,Sarkar2019,Ranjith2019a,Xing2019a,Xing2019b,Ma2020,Zangeneh2019,Zhang2020}.  Here, the eight fold degeneracy of Yb$^{3+}$ ions could be lifted to four Kramers doublets due to the combined effects of strong spin-orbit coupling (SOC) and crystal electric field (CEF). At temperatures considerably lower than the energy gap $\Delta$ between the ground state and the first excited state doublet, the magnetic properties can be described by an effective spin-$\frac{1}{2}$ local moment. The values of $\Delta$ are found to be $\sim$ 200, $\sim$400, and $\sim$ 180~K  for NaYbS$_2$\cite{baenitz2018,Joerg2019a },  NaYbO$_2$\cite{Ranjith2019,Ding2019}, and NaYbSe$_2$\cite{Ranjith2019a,Zhang2020},  respectively. Low field studies reveal a QSL ground state with gapless excitations for both NaYbS$_2$ and  NaYbO$_2$. In the case of NaYbO$_2$, upon applying the external magnetic fields, this QSL state becomes unstable and it shows magnetic long range ordering above $\mu_0H\simeq$ 2~T. This field-induced magnetic ordering of NaYbO$_2$ is well confirmed by the magnetization, heat capacity, and neutron diffraction measurements\cite{Ranjith2019,Bordelon2019}.

\begin{figure}
  \centering
  \includegraphics[width=0.5\columnwidth]{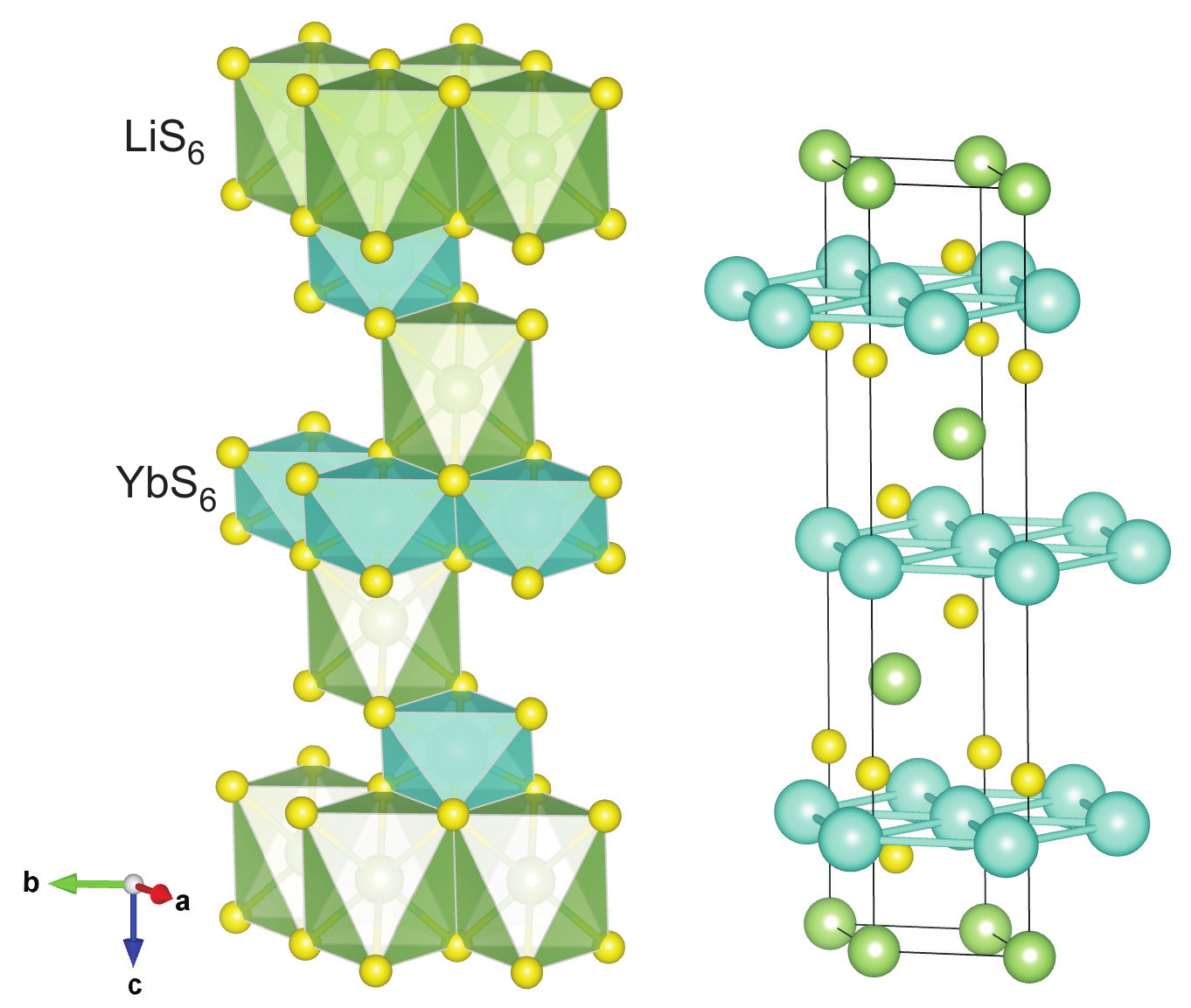}
  \caption{Crystal structure of LiYbS$_2$ composed of LiS$_6$ and YbS$_6$ octahedra. Yb-ions form triangular layers in the $ab$-plane which are separated by LiS$_6$ layers.}\label{1}
\end{figure}

Here, we investigate the ground state properties of LiYbS$_2$, an another member of the $A$Yb$X_2$ family, through magnetization and $^7$Li NMR measurements. LiYbS$_2$ crystalizes in the trigonal crystal structure with $R\bar{3}m$ space group (No:166). The crystal structure contains slightly distorted YbS$_6$ octahedra and LiS$_6$ octahedra (Fig.~\ref{1}).  The edge-shared YbS$_6$  octahedra form the perfect triangular layers of magnetic Yb-ions in the $ab$-plane, which are well separated by the non-magnetic LiS$_6$ octahedra along the crystallographic $c$ direction.
The local ground state doublet is well separated from the first excited doublet by an energy gap $\Delta\sim$ 200~K and hence the low-temperature properties can be well described by an effective spin-$\frac{1}{2}$ state~\cite{Joerg2019}.

\section{Experimental}
Polycrystalline sample of LiYbS$_2$ was synthesized with LiCl  (Alfa Aesar, 99.9\%), Li$_2$S (Alfa Aesar, 99.9\%), and YbCl$_3$ (ABCR, 99.99\%) as starting materials. All chemicals are stored and handled in an argon-filled glove-box. Prior to usage, the silica tube with glassy carbon inlayer is heated out under vacuum at 850~$^\circ$C. The starting materials were sealed in the silica tube, slowly heated to 700~$^\circ$C and held at this temperature for 2 days. After cooling down to room temperature, the ampoule was opened under ambient atmosphere and the intense greenish-yellow product extracted by washing with water several times.  An alternative synthesis route was also discussed in Ref.~\cite{cotter1994}. Phase purity of the sample was confirmed by powder X-ray diffraction experiments.

The magnetic properties were measured using a vibrating sample magnetometer (VSM) attachment in a commercial (Quantum Design) SQUID magnetometer. The nuclear magnetic resonance (NMR) experiments were carried out by applying a pulsed NMR technique on the $^7$Li nucleus (nuclear spin $I= \frac{3}{2}$) using a commercially available NMR spectrometer (Tecmag).

\section{Results and Discussions}
\begin{figure}
  \centering
  \includegraphics[width=0.8\columnwidth]{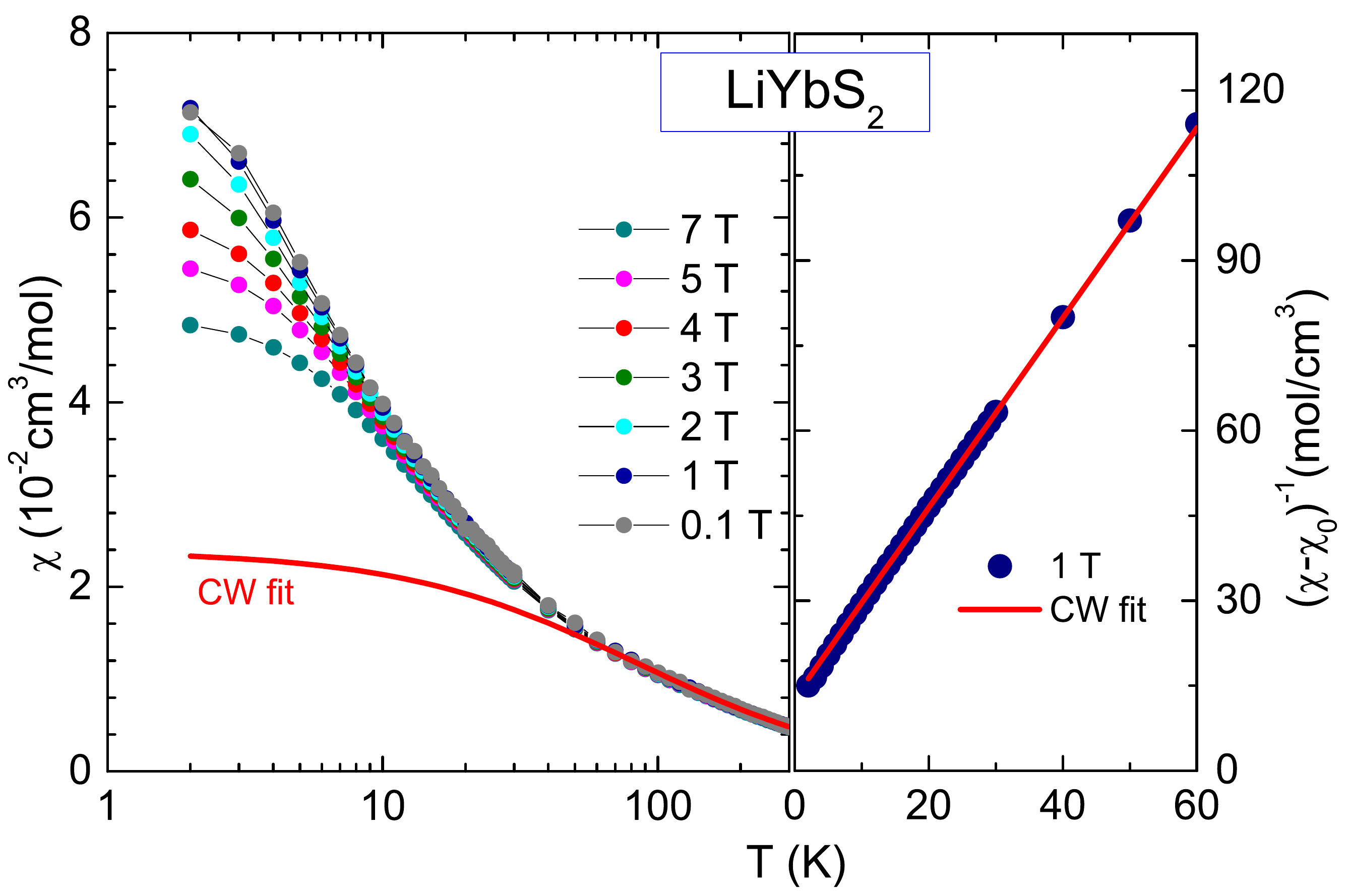}
  \caption{Left panel: temperature dependence of the magnetic susceptibility measured at different applied fields. The red solid line represents the high temperature Curie-Weiss fit. Right panel: Low temperature inverse susceptibility after the subtraction of $\chi_0$ together with the low temperature curie-Weiss fit.}\label{2}
\end{figure}
Figure~\ref{2} shows the temperature dependence of the magnetic susceptibility $\chi(T)$ of LiYbS$_2$ measured at different applied fields. $\chi(T)$ did not show any indication of magnetic long-range ordering (LRO) down to 2~K. The $\chi(T)$ data at high temperatures were fitted to $\chi(T) = \chi_0+\frac{C}{T-\theta_{\rm CW}}$, where $\chi_0$ is the temperature-independent contribution that accounts for core diamagnetism and Van-Vleck paramagnetism, while the second term is the Curie-Weiss (CW) law with the Curie constant $C=\frac{N_A\mu_{\rm eff}^2}{3k_{\rm B}}$ and the Curie-Weiss temperature $\theta_{\rm CW}$. The fitting above 100~K yields $\chi_0\simeq-5.56\times 10^{-4}$~cm$^3$/mol, $C\simeq$ 2.31~cm$^3$K/mol, and $\theta_{\rm CW}\simeq$ -110~K. The $C$ value corresponds to an effective moment $\mu_{\rm eff}$ = 4.3~$\mu_{\rm B}$ which is in good agreement with the expected value ($g\sqrt{j(j+1)}=4.54~\mu_{\rm B}$) for a Yb$^{3+}$ ion having a $^2F_{7/2}$ multiplet with $g=8/7$. However, in this high-temperature regime, the crystal electric-field (CEF) excitations of  Yb$^{3+}$ are expected to have a dominant contribution  to the Curie-Weiss temperature.

\begin{figure}
  \centering
  \includegraphics[width=0.8\columnwidth]{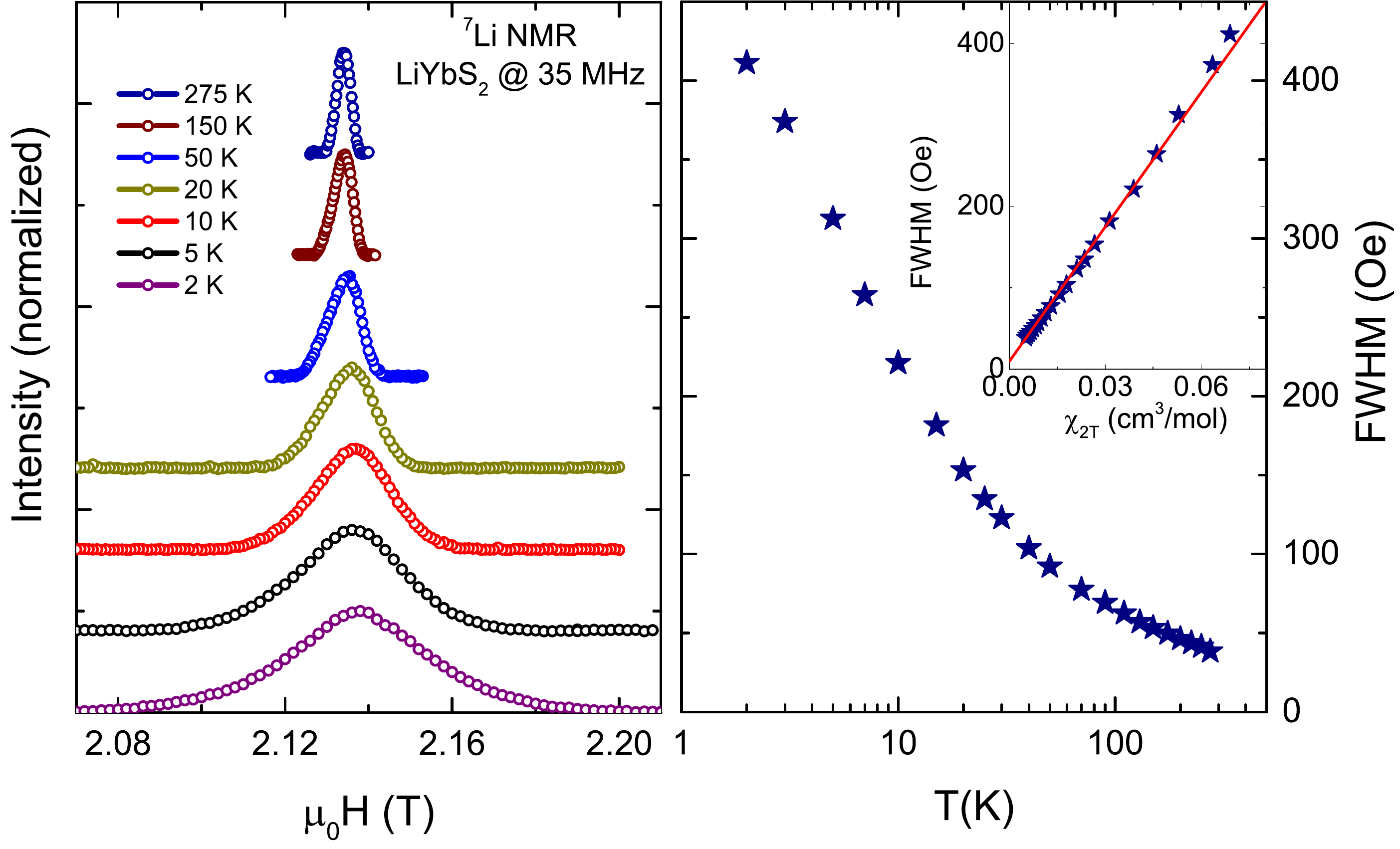}
  \caption{Left pane: $^7$Li NMR spectra measured at different temperatures. Right panel: NMR linewidth as a function of temperature. Inset shows the FWHM vs $\chi$ plot with temperature as an implicit parameter. The red line is a linear fit.}\label{3}
\end{figure}

In the low temperature regime, where the crystal electric-field excitations can be neglected,  a change of slope is observed in the inverse susceptibility associated with the Kramers ground state doublet. Below 40~K the magnetic susceptibility, after subtracting a  constant term $\chi_0=5.5~\times 10^{-3}$~cm$^3$/mol could be described by the Curie-Weiss law which is yielding $\theta_{\rm CW}\simeq$ -8~K and $\mu_{\rm eff}\simeq$ 2.3~$\mu_{\rm B}$. At low temperatures, we observed a  $\chi_0$ value, one order in magnitude larger than that of 300~K, which is due to  the van Vleck susceptibility $\chi_{\rm VV}$, arising from the CEF excitations. Similar $\chi_0$ values are also reported for NaYbO$_2$ and NaYbS$_2$ at low temperatures\cite{baenitz2018,Ranjith2019}. The obtained effective moment ($\mu_{\rm eff}=g\sqrt{S(S+1)}$) corresponds to an effective spin-$\frac{1}{2}$ state with $g\simeq$ 2.66. This value matches the powder averaged $g_{\rm av}\simeq$ 2.9 from the ESR measurements~\cite{Joerg2019}. The negative value of  $\theta_{\rm CW}$ indicates a dominant antiferromagnetic nature of exchange interactions between the Yb moments. On the mean-field level, the Curie-Weiss temperature can be expressed as $\theta_{\rm CW}=-zJS(S+1)/3k_{\rm B}$, where $z$ is the number of nearest-neighbor spins and $k_{\rm B}$ is the Boltzmann constant~\cite{schmidt:17}. For a triangular lattice system, the number of nearest neighbors is $z=6$, which yields an antiferromagnetic exchange interaction of $J/k_{\rm B}\simeq$ 5.3~K for LiYbS$_2$. Furthermore, no indications of magnetic LRO was observed down to 2~K, which suggest a possible QSL ground state for LiYbS$_2$.

We have used $^7$Li NMR as a local probe to further investigate the magnetic properties of LiYbS$_2$. $^7$Li is a spin-$\frac{3}{2}$ nucleus for which one would expect  a powder NMR spectra  with a central line together with quadrupole singularities on both sides. As shown in the left panel of Fig.~\ref{3}, our NMR experiments yield a narrow spectral line without any singularities, which is probably due to the low quadrupolar frequency or distribution of intensity of the satellites over a broad frequency range. This type of single spectral line is commonly observed in $^7$Li NMR on several low-dimensional spin systems~\cite{ranjith2015,ranjith2016}.  The NMR spectra were found to broaden monotonously upon lowering the temperature. No drastic line broadening was observed which confirms the absence of magnetic LRO down to 2~K. The NMR line width (full width half maximum, FWHM) as a function of temperature is shown in the right pane of Fig.~\ref{3}. Linewidth is increasing systematically with decreasing temperature and tracks the magnetic susceptibility. From the linear slope of the plot of linewidth vs magnetic susceptibility with temperature as an implicit parameter (Inset of Fig.~\ref{3}), one can calculate the dipolar coupling constant~\cite{ranjith2015,ranjith2018}. The obtained dipolar coupling constant $A_{\rm dip}\simeq 9\times 10^{22}$~cm$^{-3}$ is of the right order of magnitude with the expected dipolar interaction of Li nuclei with Yb moments at an average distance of 3.7~\AA.
 \section{Conclusions}
We have investigated  the Yb-based triangular lattice QSL candidate LiYbS$_2$ through magnetization and $^7$Li NMR measurements. At low temperatures, the magnetic properties can be described by an effective spin-$\frac{1}{2}$ ground state.  Magnetic susceptibility and $^7$Li NMR measurements did not show any indication of magnetic long-range order, which suggests a possible quantum spin liquid ground state for LiYbS$_2$. At low temperatures, the dominant interaction is antiferromagnetic in nature and an exchange constant of  $J/k_{\rm B}\simeq$ 5.3~K  can be extracted in the powder limit. Furthermore, from the analysis of NMR linewidth vs static susceptibility, the coupling between the Li nuclei and Yb$^{3+}$ ions was found to be mainly dipolar in nature.

 \section{Acknowledgement}

 We thank J. Sichelschmidt, B. Schmidt, H. Yasuoka, S. Luther, and H. K\"{u}hne for fruitful discussions.
Th. D. and Ph. S. thank the Deutsche Forschungsgemeinschaft for financial support within the CRC1143 framework (project B03).

%y}%


\begin{thebibliography}{13}%
\makeatletter
\providecommand \@ifxundefined [1]{%
 \@ifx{#1\undefined}
}%
\providecommand \@ifnum [1]{%
 \ifnum #1\expandafter \@firstoftwo
 \else \expandafter \@secondoftwo
 \fi
}%
\providecommand \@ifx [1]{%
 \ifx #1\expandafter \@firstoftwo
 \else \expandafter \@secondoftwo
 \fi
}%
\providecommand \natexlab [1]{#1}%
\providecommand \enquote  [1]{``#1''}%
\providecommand \bibnamefont  [1]{#1}%
\providecommand \bibfnamefont [1]{#1}%
\providecommand \citenamefont [1]{#1}%
\providecommand \href@noop [0]{\@secondoftwo}%
\providecommand \href [0]{\begingroup \@sanitize@url \@href}%
\providecommand \@href[1]{\@@startlink{#1}\@@href}%
\providecommand \@@href[1]{\endgroup#1\@@endlink}%
\providecommand \@sanitize@url [0]{\catcode `\\12\catcode `\$12\catcode
  `\&12\catcode `\#12\catcode `\^12\catcode `\_12\catcode `\%12\relax}%
\providecommand \@@startlink[1]{}%
\providecommand \@@endlink[0]{}%
\providecommand \url  [0]{\begingroup\@sanitize@url \@url }%
\providecommand \@url [1]{\endgroup\@href {#1}{\urlprefix }}%
\providecommand \urlprefix  [0]{URL }%
\providecommand \Eprint [0]{\href }%
\providecommand \doibase [0]{http://dx.doi.org/}%
\providecommand \selectlanguage [0]{\@gobble}%
\providecommand \bibinfo  [0]{\@secondoftwo}%
\providecommand \bibfield  [0]{\@secondoftwo}%
\providecommand \translation [1]{[#1]}%
\providecommand \BibitemOpen [0]{}%
\providecommand \bibitemStop [0]{}%
\providecommand \bibitemNoStop [0]{.\EOS\space}%
\providecommand \EOS [0]{\spacefactor3000\relax}%
\providecommand \BibitemShut  [1]{\csname bibitem#1\endcsname}%
\let\auto@bib@innerbib\@empty
%</preamble>
\bibitem [1]{anderson1973}%
  \BibitemOpen
  \bibfield  {author} {\bibinfo {author} {\bibfnamefont {P.~W.}\ \bibnamefont
  {Anderson}},\ }\bibfield  {title} {\enquote {\bibinfo {title} {Resonating
  valence bonds: A new kind of insulator?}}\ }\href@noop {} {\bibfield
  {journal} {\bibinfo  {journal} {Mater. Res. Bull.}\ }\textbf {\bibinfo
  {volume} {8}},\ \bibinfo {pages} {153--160} (\bibinfo {year}
  {1973})}\BibitemShut {NoStop}%
\bibitem [2]{balents2010}%
  \BibitemOpen
  \bibfield  {author} {\bibinfo {author} {\bibfnamefont {L.}~\bibnamefont
  {Balents}},\ }\bibfield  {title} {\enquote {\bibinfo {title} {Spin liquids in
  frustrated magnets},}\ }\href@noop {} {\bibfield  {journal} {\bibinfo
  {journal} {Nature}\ }\textbf {\bibinfo {volume} {464}},\ \bibinfo {pages}
  {199} (\bibinfo {year} {2010})}\BibitemShut {NoStop}%
\bibitem [3]{Zhou2017}%
  \BibitemOpen
  \bibfield  {author} {\bibinfo {author} {\bibfnamefont {Y.}~\bibnamefont
  {Zhou}}, \bibinfo {author} {\bibfnamefont {K.}~\bibnamefont {Kanoda}}, \ and\
  \bibinfo {author} {\bibfnamefont {T.-K.}\ \bibnamefont {Ng}},\ }\bibfield
  {title} {\enquote {\bibinfo {title} {Quantum spin liquid states},}\
  }\href@noop {} {\bibfield  {journal} {\bibinfo  {journal} {Rev. Mod. Phys.}\
  }\textbf {\bibinfo {volume} {89}},\ \bibinfo {pages} {025003} (\bibinfo
  {year} {2017})}\BibitemShut {NoStop}%
\bibitem [4]{Wen2019}%
  \BibitemOpen
  \bibfield  {author} {\bibinfo {author} {\bibfnamefont {J.}~\bibnamefont
  {Wen}}, \bibinfo {author} {\bibfnamefont {S.~L.}\ \bibnamefont {Yu}},
  \bibinfo {author} {\bibfnamefont {S.}~\bibnamefont {Li}}, \bibinfo {author}
  {\bibfnamefont {W.}~\bibnamefont {Yu}}, \ and\ \bibinfo {author}
  {\bibfnamefont {J-X.}\ \bibnamefont {Li}},\ }\bibfield  {title} {\enquote
  {\bibinfo {title} {Experimental identification of quantum spin liquids},}\
  }\href@noop {} {\bibfield  {journal} {\bibinfo  {journal} {npj Quantum
  Materials}\ }\textbf {\bibinfo {volume} {4}},\ \bibinfo {pages} {12}
  (\bibinfo {year} {2019})}\BibitemShut {NoStop}%
\bibitem [5]{liu2018}%
  \BibitemOpen
  \bibfield  {author} {\bibinfo {author} {\bibfnamefont {W.}~\bibnamefont
  {Liu}}, \bibinfo {author} {\bibfnamefont {Z.}~\bibnamefont {Zhang}}, \bibinfo
  {author} {\bibfnamefont {J.}~\bibnamefont {Ji}}, \bibinfo {author}
  {\bibfnamefont {Y.}~\bibnamefont {Liu}}, \bibinfo {author} {\bibfnamefont
  {J.}~\bibnamefont {Li}}, \bibinfo {author} {\bibfnamefont {X.}~\bibnamefont
  {Wang}}, \bibinfo {author} {\bibfnamefont {H.}~\bibnamefont {Lei}}, \bibinfo
  {author} {\bibfnamefont {G.}~\bibnamefont {Chen}}, \ and\ \bibinfo {author}
  {\bibfnamefont {Q.}~\bibnamefont {Zhang}},\ }\bibfield  {title} {\enquote
  {\bibinfo {title} {Rare-earth chalcogenides: A large family of triangular
  lattice spin liquid candidates},}\ }\href@noop {} {\bibfield  {journal}
  {\bibinfo  {journal} {Chin. Phys. Lett.}\ }\textbf {\bibinfo {volume} {35}},\
  \bibinfo {pages} {117501} (\bibinfo {year} {2018})}\BibitemShut {NoStop}%
\bibitem [6]{baenitz2018}%
  \BibitemOpen
  \bibfield  {author} {\bibinfo {author} {\bibfnamefont {M.}~\bibnamefont
  {Baenitz}}, \bibinfo {author} {\bibfnamefont {Ph.}\ \bibnamefont
  {Schlender}}, \bibinfo {author} {\bibfnamefont {J.}~\bibnamefont
  {Sichelschmidt}}, \bibinfo {author} {\bibfnamefont {Y.~A.}\ \bibnamefont
  {Onykiienko}}, \bibinfo {author} {\bibfnamefont {Z.}~\bibnamefont
  {Zangeneh}}, \bibinfo {author} {\bibfnamefont {K.~M.}\ \bibnamefont
  {Ranjith}}, \bibinfo {author} {\bibfnamefont {R.}~\bibnamefont {Sarkar}},
  \bibinfo {author} {\bibfnamefont {L.}~\bibnamefont {Hozoi}}, \bibinfo
  {author} {\bibfnamefont {H.~C.}\ \bibnamefont {Walker}}, \bibinfo {author}
  {\bibfnamefont {J.-C.}\ \bibnamefont {Orain}}, \bibinfo {author}
  {\bibfnamefont {H.}~\bibnamefont {Yasuoka}}, \bibinfo {author} {\bibfnamefont
  {J.}~\bibnamefont {van~den Brink}}, \bibinfo {author} {\bibfnamefont {H.~H.}\
  \bibnamefont {Klauss}}, \bibinfo {author} {\bibfnamefont {D.~S.}\
  \bibnamefont {Inosov}}, \ and\ \bibinfo {author} {\bibfnamefont {Th.}\
  \bibnamefont {Doert}},\ }\bibfield  {title} {\enquote {\bibinfo {title}
  {{NaYbS}$_2$: A planar spin-$\frac{1}{2}$ triangular-lattice magnet and
  putative spin liquid},}\ }\href@noop {}
  {\bibfield  {journal} {\bibinfo  {journal} {Phys. Rev. B}\ }\textbf {\bibinfo
  {volume} {98}},\ \bibinfo {pages} {220409(R)} (\bibinfo {year}
  {2018})}\BibitemShut {NoStop}%
\bibitem [7]{Ranjith2019}%
  \BibitemOpen
  \bibfield  {author} {\bibinfo {author} {\bibfnamefont {K.~M.}\ \bibnamefont
  {Ranjith}}, \bibinfo {author} {\bibfnamefont {D.}~\bibnamefont {Dmytriieva}},
  \bibinfo {author} {\bibfnamefont {S.}~\bibnamefont {Khim}}, \bibinfo {author}
  {\bibfnamefont {J.}~\bibnamefont {Sichelschmidt}}, \bibinfo {author}
  {\bibfnamefont {S.}~\bibnamefont {Luther}}, \bibinfo {author} {\bibfnamefont
  {D.}~\bibnamefont {Ehlers}}, \bibinfo {author} {\bibfnamefont
  {H.}~\bibnamefont {Yasuoka}}, \bibinfo {author} {\bibfnamefont
  {J.}~\bibnamefont {Wosnitza}}, \bibinfo {author} {\bibfnamefont {A.~A.}\
  \bibnamefont {Tsirlin}}, \bibinfo {author} {\bibfnamefont {H.}~\bibnamefont
  {K\"uhne}}, \ and\ \bibinfo {author} {\bibfnamefont {M.}~\bibnamefont
  {Baenitz}},\ }\bibfield  {title} {\enquote {\bibinfo {title} {Field-induced
  instability of the quantum spin liquid ground state in the
  ${J}_{\mathrm{eff}}=\frac{1}{2}$ triangular-lattice compound {NaYbO}$_2$},}\
  }\href@noop {} {\bibfield  {journal} {\bibinfo
  {journal} {Phys. Rev. B}\ }\textbf {\bibinfo {volume} {99}},\ \bibinfo
  {pages} {180401(R)} (\bibinfo {year} {2019})}\BibitemShut {NoStop}%
\bibitem [8]{Ding2019}%
  \BibitemOpen
  \bibfield  {author} {\bibinfo {author} {\bibfnamefont {L.}~\bibnamefont
  {Ding}}, \bibinfo {author} {\bibfnamefont {P.}~\bibnamefont {Manuel}},
  \bibinfo {author} {\bibfnamefont {S.}~\bibnamefont {Bachus}}, \bibinfo
  {author} {\bibfnamefont {F.}~\bibnamefont {Gru{\ss}ler}}, \bibinfo {author}
  {\bibfnamefont {P.}~\bibnamefont {Gegenwart}}, \bibinfo {author}
  {\bibfnamefont {J.}~\bibnamefont {Singleton}}, \bibinfo {author}
  {\bibfnamefont {R.D.}\ \bibnamefont {Johnson}}, \bibinfo {author}
  {\bibfnamefont {H.C.}\ \bibnamefont {Walker}}, \bibinfo {author}
  {\bibfnamefont {D.T.}\ \bibnamefont {Adroja}}, \bibinfo {author}
  {\bibfnamefont {A.D.}\ \bibnamefont {Hillier}}, \ and\ \bibinfo {author}
  {\bibfnamefont {A.A.}\ \bibnamefont {Tsirlin}},\ }\bibfield  {title}
  {\enquote {\bibinfo {title} {Gapless spin-liquid state in the structurally
  disorder-free triangular antiferromagnet {NaYbO$_2$}},}\ }\href@noop {}
  {\bibfield  {journal} {\bibinfo  {journal} {arXiv:1901.07810}\ } (\bibinfo
  {year} {2019})}\BibitemShut {NoStop}%
\bibitem [9]{Bordelon2019}%
  \BibitemOpen
  \bibfield  {author} {\bibinfo {author} {\bibfnamefont {M.~M.}\ \bibnamefont
  {Bordelon}}, \bibinfo {author} {\bibfnamefont {E.}~\bibnamefont {Kenney}},
  \bibinfo {author} {\bibfnamefont {C.}~\bibnamefont {Liu}}, \bibinfo {author}
  {\bibfnamefont {T.}~\bibnamefont {Hogan}}, \bibinfo {author} {\bibfnamefont
  {L.}~\bibnamefont {Posthuma}}, \bibinfo {author} {\bibfnamefont
  {M.}~\bibnamefont {Kavand}}, \bibinfo {author} {\bibfnamefont
  {Y.}~\bibnamefont {Lyu}}, \bibinfo {author} {\bibfnamefont {M.}~\bibnamefont
  {Sherwin}}, \bibinfo {author} {\bibfnamefont {N.~P.}\ \bibnamefont {Butch}},
  \bibinfo {author} {\bibfnamefont {C.}~\bibnamefont {Brown}}, \bibinfo
  {author} {\bibfnamefont {M.~J.}\ \bibnamefont {Graf}}, \bibinfo {author}
  {\bibfnamefont {L.}~\bibnamefont {Balents}}, \ and\ \bibinfo {author}
  {\bibfnamefont {S.~D.}\ \bibnamefont {Wilson}},\ }\bibfield  {title}
  {\enquote {\bibinfo {title} {Field-tunable quantum disordered ground state in
  the triangular-lattice antiferromagnet {NaYbO$_2$}},}\ }\href@noop {} {\bibfield  {journal} {\bibinfo
  {journal} {Nat. Phys.}\ } (\bibinfo {year} {2019})}\BibitemShut {NoStop}%
\bibitem [10]{Sarkar2019}%
  \BibitemOpen
  \bibfield  {author} {\bibinfo {author} {\bibfnamefont {R.}~\bibnamefont
  {Sarkar}}, \bibinfo {author} {\bibfnamefont {Ph.}\ \bibnamefont {Schlender}},
  \bibinfo {author} {\bibfnamefont {V.}~\bibnamefont {Grinenko}}, \bibinfo
  {author} {\bibfnamefont {E.}~\bibnamefont {Haeussler}}, \bibinfo {author}
  {\bibfnamefont {Peter~J.}\ \bibnamefont {Baker}}, \bibinfo {author}
  {\bibfnamefont {Th.}\ \bibnamefont {Doert}}, \ and\ \bibinfo {author}
  {\bibfnamefont {H.-H.}\ \bibnamefont {Klauss}},\ }\bibfield  {title}
  {\enquote {\bibinfo {title} {Quantum spin liquid ground state in the disorder
  free triangular lattice {NaYbO$_2$}},}\ }\href@noop {} {\bibinfo  {journal}
  {arXiv:1911.08036}\ }\BibitemShut {NoStop}%
\bibitem [11]{Ranjith2019a}%
  \BibitemOpen
\bibfield  {journal} {  }\bibfield  {author} {\bibinfo {author} {\bibfnamefont
  {K.~M.}\ \bibnamefont {Ranjith}}, \bibinfo {author} {\bibfnamefont
  {S.}~\bibnamefont {Luther}}, \bibinfo {author} {\bibfnamefont
  {T.}~\bibnamefont {Reimann}}, \bibinfo {author} {\bibfnamefont
  {B.}~\bibnamefont {Schmidt}}, \bibinfo {author} {\bibfnamefont {Ph.}\
  \bibnamefont {Schlender}}, \bibinfo {author} {\bibfnamefont {J.}~\bibnamefont
  {Sichelschmidt}}, \bibinfo {author} {\bibfnamefont {H.}~\bibnamefont
  {Yasuoka}}, \bibinfo {author} {\bibfnamefont {A.~M.}\ \bibnamefont
  {Strydom}}, \bibinfo {author} {\bibfnamefont {Y.}~\bibnamefont {Skourski}},
  \bibinfo {author} {\bibfnamefont {J.}~\bibnamefont {Wosnitza}}, \bibinfo
  {author} {\bibfnamefont {H.}~\bibnamefont {K\"uhne}}, \bibinfo {author}
  {\bibfnamefont {Th.}\ \bibnamefont {Doert}}, \ and\ \bibinfo {author}
  {\bibfnamefont {M.}~\bibnamefont {Baenitz}},\ }\bibfield  {title} {\enquote
  {\bibinfo {title} {Anisotropic field-induced ordering in the
  triangular-lattice quantum spin liquid ${\mathrm{NaYbSe}}_{2}$},}\
  }\href@noop {} {\bibfield  {journal} {\bibinfo
  {journal} {Phys. Rev. B}\ }\textbf {\bibinfo {volume} {100}},\ \bibinfo
  {pages} {224417} (\bibinfo {year} {2019})}\BibitemShut {NoStop}%
\bibitem [12]{Xing2019a}%
  \BibitemOpen
  \bibfield  {author} {\bibinfo {author} {\bibfnamefont {J.}\ \bibnamefont
  {Xing}}, \bibinfo {author} {\bibfnamefont {L.~D.}\ \bibnamefont
  {Sanjeewa}}, \bibinfo {author} {\bibfnamefont {J.}\ \bibnamefont {Kim}},
  \bibinfo {author} {\bibfnamefont {G.~R.}\ \bibnamefont {Stewart}}, \bibinfo
  {author} {\bibfnamefont {A.}\ \bibnamefont {Podlesnyak}}, \ and\ \bibinfo
  {author} {\bibfnamefont {A.~S.}\ \bibnamefont {Sefat}},\ }\bibfield
  {title} {\enquote {\bibinfo {title} {Field-induced magnetic transition and
  spin fluctuations in the quantum spin-liquid candidate
  ${\mathrm{CsYbSe}}_{2}$},}\ }\href@noop {}
  {\bibfield  {journal} {\bibinfo  {journal} {Phys. Rev. B}\ }\textbf {\bibinfo
  {volume} {100}},\ \bibinfo {pages} {220407} (\bibinfo {year}
  {2019})}\BibitemShut {NoStop}%
\bibitem [13]{Xing2019b}%
  \BibitemOpen
  \bibfield  {author} {\bibinfo {author} {\bibfnamefont {J.}\ \bibnamefont
  {Xing}}, \bibinfo {author} {\bibfnamefont {L.~D}\ \bibnamefont
  {Sanjeewa}}, \bibinfo {author} {\bibfnamefont {J.}\ \bibnamefont {Kim}},
  \bibinfo {author} {\bibfnamefont {G.~R}~\bibnamefont {Stewart}}, \bibinfo
  {author} {\bibfnamefont {M-H.}\ \bibnamefont {Du}}, \bibinfo {author}
  {\bibfnamefont {F.~A}\ \bibnamefont {Reboredo}}, \bibinfo {author}
  {\bibfnamefont {R.}\ \bibnamefont {Custelcean}}, \ and\ \bibinfo {author}
  {\bibfnamefont {A.~S}\ \bibnamefont {Sefat}},\ }\bibfield  {title}
  {\enquote {\bibinfo {title} {Crystal synthesis and frustrated magnetism in
  triangular lattice {CsReSe$_2$ (Re= La-Lu)}: Quantum spin liquid candidates
  {CsCeSe$_2$ and CsYbSe$_2$}},}\ }\href@noop {} {\bibfield  {journal} {\bibinfo
  {journal} {ACS Materials Letters}\ } (\bibinfo {year}
  {2019}{\natexlab{b}})}\BibitemShut {NoStop}%
\bibitem [14]{Ma2020}%
  \BibitemOpen
\bibfield  {journal} {  }\bibfield  {author} {\bibinfo {author} {\bibfnamefont
  {J.}~\bibnamefont {Ma}}, \bibinfo {author} {\bibfnamefont {J.}~\bibnamefont
  {Li}}, \bibinfo {author} {\bibfnamefont {Y.~H.}\ \bibnamefont {Gao}},
  \bibinfo {author} {\bibfnamefont {C.}~\bibnamefont {Liu}}, \bibinfo {author}
  {\bibfnamefont {Q.}~\bibnamefont {Ren}}, \bibinfo {author} {\bibfnamefont
  {Z.}~\bibnamefont {Zhang}}, \bibinfo {author} {\bibfnamefont
  {Z.}~\bibnamefont {Wang}}, \bibinfo {author} {\bibfnamefont {R.}~\bibnamefont
  {Chen}}, \bibinfo {author} {\bibfnamefont {J.}~\bibnamefont {Embs}}, \bibinfo
  {author} {\bibfnamefont {E.}~\bibnamefont {Feng}}, \bibinfo {author}
  {\bibfnamefont {F.}~\bibnamefont {Zhu}}, \bibinfo {author} {\bibfnamefont
  {Q.}~\bibnamefont {Huang}}, \bibinfo {author} {\bibfnamefont
  {Z.}~\bibnamefont {Xiang}}, \bibinfo {author} {\bibfnamefont
  {L.}~\bibnamefont {Chen}}, \bibinfo {author} {\bibfnamefont {E.~S.}\
  \bibnamefont {Choi}}, \bibinfo {author} {\bibfnamefont {Z.}~\bibnamefont
  {Qu}}, \bibinfo {author} {\bibfnamefont {L.}~\bibnamefont {Li}}, \bibinfo
  {author} {\bibfnamefont {J.}~\bibnamefont {Wang}}, \bibinfo {author}
  {\bibfnamefont {H.}~\bibnamefont {Zhou}}, \bibinfo {author} {\bibfnamefont
  {Y.}~\bibnamefont {Su}}, \bibinfo {author} {\bibfnamefont {X.}~\bibnamefont
  {Wang}}, \bibinfo {author} {\bibfnamefont {Q.}~\bibnamefont {Zhang}}, \ and\
  \bibinfo {author} {\bibfnamefont {G.}~\bibnamefont {Chen}},\ }\bibfield
  {title} {\enquote {\bibinfo {title} {Spin-orbit-coupled triangular-lattice
  spin liquid in rare-earth chalcogenides},}\ }\href@noop {} {\bibinfo
  {journal} {arXiv:2002.09224}\ }\BibitemShut {NoStop}%
\bibitem [15]{Zangeneh2019}%
  \BibitemOpen
\bibfield  {journal} {  }\bibfield  {author} {\bibinfo {author} {\bibfnamefont
  {Z.}~\bibnamefont {Zangeneh}}, \bibinfo {author} {\bibfnamefont
  {S.}~\bibnamefont {Avdoshenko}}, \bibinfo {author} {\bibfnamefont
  {J.}~\bibnamefont {van~den Brink}}, \ and\ \bibinfo {author} {\bibfnamefont
  {L.}~\bibnamefont {Hozoi}},\ }\bibfield  {title} {\enquote {\bibinfo {title}
  {Single-site magnetic anisotropy governed by interlayer cation charge
  imbalance in triangular-lattice $A\mathrm{Yb}{X}_{2}$},}\ }\href@noop {}{\bibfield  {journal} {\bibinfo  {journal}
  {Phys. Rev. B}\ }\textbf {\bibinfo {volume} {100}},\ \bibinfo {pages}
  {174436} (\bibinfo {year} {2019})}\BibitemShut {NoStop}%
\bibitem [16]{Zhang2020}%
  \BibitemOpen
\bibfield  {journal} {  }\bibfield  {author} {\bibinfo {author} {\bibfnamefont
  {Z.}~\bibnamefont {Zhang}}, \bibinfo {author} {\bibfnamefont
  {X.}~\bibnamefont {Ma}}, \bibinfo {author} {\bibfnamefont {J.}~\bibnamefont
  {Li}}, \bibinfo {author} {\bibfnamefont {G.}~\bibnamefont {Wang}}, \bibinfo
  {author} {\bibfnamefont {D.~T.}\ \bibnamefont {Adroja}}, \bibinfo {author}
  {\bibfnamefont {T.~G.}\ \bibnamefont {Perring}}, \bibinfo {author}
  {\bibfnamefont {W.}~\bibnamefont {Liu}}, \bibinfo {author} {\bibfnamefont
  {F.}~\bibnamefont {Jin}}, \bibinfo {author} {\bibfnamefont {J.}~\bibnamefont
  {Ji}}, \bibinfo {author} {\bibfnamefont {Y.}~\bibnamefont {Wang}}, \bibinfo
  {author} {\bibfnamefont {X.}~\bibnamefont {Wang}}, \bibinfo {author}
  {\bibfnamefont {J.}~\bibnamefont {Ma}}, \ and\ \bibinfo {author}
  {\bibfnamefont {Q.}~\bibnamefont {Zhang}},\ }\bibfield  {title} {\enquote
  {\bibinfo {title} {Crystalline electric-field excitations in quantum spin
  liquids candidate {NaYbSe$_2$}},}\ }\href@noop {} {\bibinfo  {journal}
  {arXiv:2002.04772}\ }\BibitemShut {NoStop}%
\bibitem [17]{Joerg2019a}%
  \BibitemOpen
  \bibfield  {author} {\bibinfo {author} {\bibfnamefont {J.}~\bibnamefont
  {Sichelschmidt}}, \bibinfo {author} {\bibfnamefont {Ph.}\ \bibnamefont
  {Schlender}}, \bibinfo {author} {\bibfnamefont {B.}~\bibnamefont {Schmidt}},
  \bibinfo {author} {\bibfnamefont {M.}~\bibnamefont {Baenitz}}, \ and\
  \bibinfo {author} {\bibfnamefont {Th.}\ \bibnamefont {Doert}},\ }\bibfield
  {title} {\enquote {\bibinfo {title} {Electron spin resonance on the spin-1/2
  triangular magnet {NaYbS}$_2$},}\ }\href@noop {}
  {\bibfield  {journal} {\bibinfo  {journal} {J. Phys. Condens. Matter}\
  }\textbf {\bibinfo {volume} {31}},\ \bibinfo {pages} {205601} (\bibinfo
  {year} {2019})}\BibitemShut {NoStop}%
\bibitem [18]{Joerg2019}%
  \BibitemOpen
  \bibfield  {author} {\bibinfo {author} {\bibfnamefont {J.}~\bibnamefont
  {Sichelschmidt}}, \bibinfo {author} {\bibfnamefont {B.}~\bibnamefont
  {Schmidt}}, \bibinfo {author} {\bibfnamefont {Ph.}\ \bibnamefont
  {Schlender}}, \bibinfo {author} {\bibfnamefont {VS.}\ \bibnamefont {Khim}},
  \bibinfo {author} {\bibfnamefont {Th.}\ \bibnamefont {Doert}}, \ and\
  \bibinfo {author} {\bibfnamefont {M.}~\bibnamefont {Baenitz}},\ }\bibfield
  {title} {\enquote {\bibinfo {title} {Effective spin-1/2 moments on a
  {Yb$^{3+}$} triangular lattice: an {ESR} study},}\ }\href@noop {} {\bibinfo
  {journal} {arXiv:1912.01868}\ }\BibitemShut {NoStop}%
\bibitem [19]{cotter1994}%
  \BibitemOpen
  \bibfield  {author} {\bibinfo {author} {\bibfnamefont {J. P.}~\bibnamefont
  {Cotter}}, \bibinfo {author} {\bibfnamefont {J. C.}\ \bibnamefont
  {Fitzmaurice}}, \ and\
  \bibinfo {author} {\bibfnamefont {I. V.}\ \bibnamefont {Parkin}},\ }\bibfield
  {title} {\enquote {\bibinfo {title} {New routes to alkali-metal–rare-earth-metal sulfides},}\ }\href@noop {}
  {\bibfield  {journal} {\bibinfo  {journal} {J. Mat. Chem}\
  }\textbf {\bibinfo {volume} {4}},\ \bibinfo {pages} {1603} (\bibinfo
  {year} {1994})}\BibitemShut {NoStop}%

\bibitem [20]{schmidt:17}%
  \BibitemOpen
  \bibfield  {author} {\bibinfo {author} {\bibfnamefont {B.}~\bibnamefont
  {Schmidt}}\ and\ \bibinfo {author} {\bibfnamefont {P.}~\bibnamefont
  {Thalmeier}},\ }\bibfield  {title} {\enquote {\bibinfo {title} {Frustrated
  two dimensional quantum magnets},}\ }\href@noop {}{\bibfield  {journal} {\bibinfo  {journal}
  {Phys. Rep.}\ }\textbf {\bibinfo {volume} {703}},\ \bibinfo {pages} {1--59}
  (\bibinfo {year} {2017})}\BibitemShut {NoStop}%
\bibitem [21]{ranjith2015}%
  \BibitemOpen
  \bibfield  {author} {\bibinfo {author} {\bibfnamefont {K.~M.}\ \bibnamefont
  {Ranjith}}, \bibinfo {author} {\bibfnamefont {M.}~\bibnamefont {Majumder}},
  \bibinfo {author} {\bibfnamefont {M.}~\bibnamefont {Baenitz}}, \bibinfo
  {author} {\bibfnamefont {A.~A.}\ \bibnamefont {Tsirlin}}, \ and\ \bibinfo
  {author} {\bibfnamefont {R.}~\bibnamefont {Nath}},\ }\bibfield  {title}
  {\enquote {\bibinfo {title} {Frustrated three-dimensional antiferromagnet
  ${\text{Li}}_{2}{\text{CuW}}_{2}{\text{O}}_{8}$: $^{7}\mathrm{Li}$ nmr and
  the effect of nonmagnetic dilution},}\ }\href@noop {}{\bibfield  {journal} {\bibinfo  {journal} {Phys.
  Rev. B}\ }\textbf {\bibinfo {volume} {92}},\ \bibinfo {pages} {024422}
  (\bibinfo {year} {2015})}\BibitemShut {NoStop}
\bibitem [22]{ranjith2016}%
  \BibitemOpen
  \bibfield  {author} {\bibinfo {author} {\bibfnamefont {K.~M.}\ \bibnamefont
  {Ranjith}}, \bibinfo {author} {\bibfnamefont {R.}~\bibnamefont {Nath}},
  \bibinfo {author} {\bibfnamefont {M.}~\bibnamefont {Majumder}}, \bibinfo
  {author} {\bibfnamefont {D.}~\bibnamefont {Kasinathan}}, \bibinfo {author}
  {\bibfnamefont {M.}~\bibnamefont {Skoulatos}}, \bibinfo {author}
  {\bibfnamefont {L.}~\bibnamefont {Keller}}, \bibinfo {author} {\bibfnamefont
  {Y.}~\bibnamefont {Skourski}}, \bibinfo {author} {\bibfnamefont
  {M.}~\bibnamefont {Baenitz}}, \ and\ \bibinfo {author} {\bibfnamefont
  {A.~A.}\ \bibnamefont {Tsirlin}},\ }\bibfield  {title} {\enquote {\bibinfo
  {title} {Commensurate and incommensurate magnetic order in spin-1 chains
  stacked on the triangular lattice in {Li$_2$NiW$_2$O$_8$}},}\ }\href@noop {}
  {\bibfield  {journal} {\bibinfo  {journal} {Phys. Rev. B}\ }\textbf {\bibinfo
  {volume} {94}},\ \bibinfo {pages} {014415} (\bibinfo {year}
  {2016})}\BibitemShut {NoStop}%
\bibitem [23]{ranjith2018}%
  \BibitemOpen
  \bibfield  {author} {\bibinfo {author} {\bibfnamefont {K.~M.}\ \bibnamefont
  {Ranjith}}, \bibinfo {author} {\bibfnamefont {C.}~\bibnamefont {Klein}},
  \bibinfo {author} {\bibfnamefont {A.~A.}\ \bibnamefont {Tsirlin}}, \bibinfo
  {author} {\bibfnamefont {H.}~\bibnamefont {Rosner}}, \bibinfo {author}
  {\bibfnamefont {C.}~\bibnamefont {Krellner}}, \ and\ \bibinfo {author}
  {\bibfnamefont {M.}~\bibnamefont {Baenitz}},\ }\bibfield  {title} {\enquote
  {\bibinfo {title} {Magnetic resonance as a local probe for kagom{\'e}
  magnetism in barlowite {Cu$_4$(OH)$_6$FBr}},}\ }\href@noop {} {\bibfield
  {journal} {\bibinfo  {journal} {Sci. Rep.}\ }\textbf {\bibinfo {volume}
  {8}},\ \bibinfo {pages} {10851} (\bibinfo {year} {2018})}\BibitemShut
  {NoStop}%
\end{thebibliography}
\end{document}